\documentclass[11pt,oneside]{article}
\usepackage[english,activeacute]{babel}
\usepackage{amssymb}
\usepackage{layout}
\usepackage[Conny]{fncychap}
\usepackage[dvips]{color}
\usepackage[dvips]{graphicx}
\usepackage[dvips]{graphics}
\usepackage[reqno]{amsmath}
\usepackage{harvard}
\usepackage{amsthm}
\usepackage{amsmath,amscd}
\usepackage{euscript}
\usepackage{pst-all}
\usepackage{pstcol}
\usepackage{enumerate}
\usepackage[dvips]{color,graphicx}
\newtheorem{defin}{\textbf{Proposition}}[section]

\theoremstyle{definition}

 \theoremstyle{remark}

\numberwithin{equation}{section}
\usepackage{array}
\usepackage{layout}

\title{\bf A weakly informative prior for Bayesian dynamic model selection with applications in fMRI}

\author{Jairo A. F\'uquene Pati\~no$^{a}$, Brenda Betancourt$^{b}$, Jo\~ao B. M. Pereira$^{c}$\\
\\
\small $^{a}$ Department of Statistics, University of Warwick, UK.
\\
\small $^{b}$ Department of Statistical Science, Duke University, USA.
\\ 
\small $^{c}$ Instituto de Matem\'atica, Universidade Federal do Rio de Janeiro, Brazil.
}

\begin{document}

\date{}

\maketitle

\begin{abstract}

In recent years, Bayesian statistics methods in neuroscience have
been showing important advances. In particular, detection of brain
signals for studying the complexity of the brain is an active area of
research. Functional magnetic resonance imagining (fMRI) is
an important tool to determine which parts of the brain are
activated by different types of physical behavior. According to recent 
results there is evidence that the values of the connectivity brain signal 
parameters are close to zero and due to the
nature of time series fMRI data with high frequency behavior,
Bayesian dynamic models for identifying sparsity are indeed
far-reaching.  We propose a multivariate Bayesian
dynamic approach for model selection and shrinkage estimation of
the connectivity parameters. We describe the coupling
or lead-lag between any pair of regions by using mixture priors
for the connectivity parameters and propose a new weakly informative
default prior for the state variances. This framework produces one-step-ahead
proper posterior predictive results and induces shrinkage and
robustness suitable for fMRI data in the presence of sparsity. To
explore the performance of the proposed methodology we present simulation
studies and an application to functional
magnetic resonance imaging data.

\end{abstract}

\section{Introduction}

\enlargethispage{5cm}

Technology in neuroscience has shown important advances over the last two decades. In particular, functional magnetic resonance imaging (fMRI) has become a powerful technique for studying the complexity of the brain and statistical analysis of this data is an active area of research (\citeasnoun{Friston}, \citeasnoun{bookfMRI} and \citeasnoun{Eddy}). One of the objectives of analyzing fMRI data is to determine which parts of the brain are activated by different types of physical sensations or activities. The signal measured in fMRI experiments is called blood-oxygen-level dependent (BOLD) response which is a consequence of
hemodynamic changes, including local changes in the blood flow, volume and oxygenation level, occurring within a few seconds of changes in neuronal activity
induced by external stimuli. This underlying hemodynamic changes associated with neural activity are commonly referred to as the hemodynamic response function (HRF).

\clearpage

\begin{figure}[htbp]
\begin{center}
\includegraphics[scale=0.7,keepaspectratio]{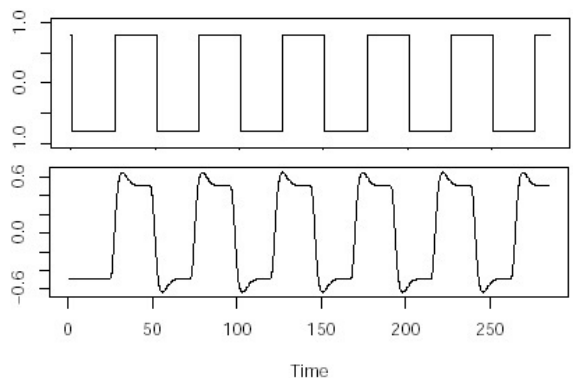}
\caption{Stimulus and hemodynamic response function of fMRI data experiment.}
\label{fig:stimulus}
\end{center}
\end{figure}

A typical BOLD response denoted by $x(t)$, where $t$ corresponds to time, usually occurs between 3 to 10 seconds after the application of the stimulus, $s(t)$, and reaches
its peak approximately after 6 seconds \cite**{Banish}.
To generate the BOLD signal, the stimulus function is convolved with a hemodynamic
response function (HRF), denoted by $h(t)$, as follows:

\begin{equation}\label{hfr}
x(t)=\int_{0}^{t}h(u)s(t-u)du,
\end{equation}

where $s(t)$ takes the value 1 when the stimulus is ON and 0 when
the stimulus is OFF, and $u$ indexes the peristimulus
time (PST) (time of neuronal firing in relation to an external
stimulus). A BOLD response can be generated based on
the time of the experiment, a microtime resolution and the ON/OFF
sets where the role of the microtime resolution is to ensure a high
precision convolution with the specific HRF. Figure \ref{fig:stimulus} displays the
stimulus and the respective hemodynamic response function of the
experiment that we present in Section \ref{sec:apply}. 


A common approach is to estimate the magnitude of the BOLD signal by considering a general linear model described as
\begin{equation}\label{form1}
y_{i,t}=\alpha_{i}+\theta_{i} x_{i,t}+ \nu_{i,t},
\end{equation}
where $y_{i,t}$ corresponds to the fMRI response at time $t$ at
voxel $i$ (a voxel is a value on a regular grid in
a three-dimensional space analogous to a pixel in a two-dimensional
space), and $\nu_{i,t}$ corresponds to the measurement noise.
The coefficient $\theta_{i}$ measures the
``activation" at voxel $i$ and represents the magnitude of the
BOLD signal at time $t$ at voxel $i$, $x_{i,t}$. Lastly,
$\alpha_{i}$ represents the baseline trend at voxel $i$,
i.e., the base effect on the fMRI response when the
effect of the BOLD signal is zero. More complex
models assume that $\alpha_i$ varies on time representing the
contribution of nuisance covariates at time $t$, for example,
periodic fluctuations due to heart rate, respiration, and head
motion. Usually, a linear smoother is used to detrend the fMRI
data. In equation (\ref{form1}), the ``activation" coefficient
$\theta_{i}$ is assumed to be invariant over time and is estimated
using maximum likelihood estimation.  However,
research suggests this parameter may vary over time. Many studies
report the detection of a strong fMRI activation in the beginning
of the experiment that becomes weaker later on. Also, it is known
that brain areas may interact with one another depending on the
context (see \citeasnoun{ringo}). For these reasons, time-varying
``activation" as well as the dependence between brain areas should
be considered in the modeling framework. A second approach that takes both features into
account, is a time-varying parameter regression which allows
time-varying connectivity between two brain regions. Here, 
differently from what it is assumed in equation
(\ref{form1}), a time series $y_{1,t}$ associated with a brain
region is regressed on a time series $y_{2,t}$ associated with
another brain region as follows:
\begin{align}\label{mmodelt}
y_{1,t}&=\theta_{t}y_{2,t} + \nu_{t},\\ \theta_{t}&=\theta_{t-1} +\omega_{t}, \notag
\end{align}
where $\theta_{t}$ measures the dynamic effective connectivity between the two brain regions, and $\nu_{t}$ and $\omega_{t}$ correspond to independent white noises \cite{Buchel}.\\

Other time-varying approaches that consider
dependence among brain areas are proposed by \citeasnoun{ombao}
and \citeasnoun{ringo}. Specifically, these authors explore
time-varying approaches for three brain regions. To study the
connectivity among them, \citename{ombao} proposed the following space-state
model:
\begin{align}\label{unoombao}
\begin{pmatrix}
  y_{1,t}\\
  y_{2,t}\\
  y_{3,t}
\end{pmatrix}
&=\begin{pmatrix}
  \alpha_{1} \\
  \alpha_{2} \\
  \alpha_{3}
\end{pmatrix}
+\begin{pmatrix}
  x_{1,t} & 0 & 0\\
  0 & x_{2,t} & 0\\
  0 & 0 & x_{3,t}
\end{pmatrix}
\begin{pmatrix}
  \theta_{1,t}\\
  \theta_{2,t}\\
  \theta_{3,t}
\end{pmatrix}
+
\begin{pmatrix}
  \nu_{1,t}\\
  \nu_{2,t}\\
  \nu_{3,t}
\end{pmatrix},
\end{align}
\begin{align}\label{dosombao}
\begin{pmatrix}
  \theta_{1,t}\\
  \theta_{2,t}\\
  \theta_{3,t}
\end{pmatrix}
&=\begin{pmatrix}
  \phi_{11}x_{1,t-1} & \phi_{12}x_{2,t-1} & \phi_{13}x_{3,t-1} \\
  \phi_{21}x_{1,t-1} & \phi_{22}x_{2,t-1} & \phi_{23}x_{3,t-1} \\
  \phi_{31}x_{1,t-1} & \phi_{32}x_{2,t-1} &
  \phi_{33}x_{3,t-1}
\end{pmatrix}
\begin{pmatrix}
  \theta_{1,t-1}\\
  \theta_{2,t-1}\\
  \theta_{3,t-1}
\end{pmatrix}
+
\begin{pmatrix}
  \omega_{1,t}\\
  \omega_{2,t}\\
  \omega_{3,t}
\end{pmatrix},
\end{align}
where $x_{i,t}$ is the hemodynamic response function at time $t$. The noise vectors $\boldsymbol{\omega}_{t}$ and
$\boldsymbol{\nu}_{t}$ are assumed to be Gaussian and independent,
\begin{align*}
\boldsymbol{\omega}_{t}&\sim N_{3}\left(0,\begin{pmatrix}
  \sigma^{2}_{\omega_{1}} & 0 & 0 \\
  0 & \sigma^{2}_{\omega_{2}} & 0 \\
  0 & 0 & \sigma^{2}_{\omega_{3}}
\end{pmatrix}\right),
& \boldsymbol{\nu}_{t}&\sim N_{3}\left(0,\begin{pmatrix}
  \sigma^{2}_{\nu_{1}} & 0 & 0 \\
  0 & \sigma^{2}_{\nu_{2}} & 0 \\
  0 & 0 & \sigma^{2}_{\nu_{3}}
\end{pmatrix}\right).
\end{align*}
The model is
determined by state parameters
$\boldsymbol{\theta}_{t}=\left\{\theta_{1,t},\theta_{2,t},
\theta_{3,t}\right\}$ linearly associated with observations
$\boldsymbol{y}_{t}=\left\{y_{1,t},y_{2,t},y_{3,t} \right\}$,
respectively. Note that equation (\ref{unoombao})
has the same structure as equation
(\ref{form1}), this equation is commonly known as the observation
equation. Equation (\ref{dosombao}) is called the
state equation and describes the dynamic of the states in a
first-order vector autoregressive model conditional on the
parameters, $\phi_{ij}$, $i=1,\dots,3$, $j=1,\dots,3$, where 
$\phi_{ij}$ represent the connectivity between the brain regions $i$ and $j$. The
initial state vector $\boldsymbol{\theta}_{0}$ is assumed to
follow a Normal distribution,
$N_{3}(\boldsymbol{\mu}_{0},\Sigma_{0})$, and is also assumed to be
independent from the noise vectors $\boldsymbol{\omega}_{t}$ and
$\boldsymbol{\nu_{t}}$.\\

\citeasnoun{ombao} use the Expectation-Maximization (EM) algorithm
to estimate all the parameters of the model \cite{Shumway}. In turn, \citeasnoun{ringo}
extended the previous proposal using the Bayesian paradigm as well
as exploring different models. In the Bayesian setting, prior information can be incorporated in the
modeling and the parameters are then estimated based on both the
data and the prior information. These proposals are
very significant as they open the door to the use of dynamic
models for investigating connectivity among brain signals.
However, some questions are left unadressed. According to
\citename{ombao} and \citename{ringo}, the values of the
connectivity parameters are close to zero. Therefore,
a natural question arises: do we need to induce some shrinkage on
the activation parameters $\boldsymbol\theta_{t}$ and connectivity
parameters $\phi_{ij}$? When is a connectivity parameter really
equal to zero? In other words, what is the probability of having a
connectivity parameter equal to zero? In addition,
the authors only take into account some of the
possible models for model selection purposes. In fact, in both approaches the
connectivity issue is only considered as an estimation problem
instead of an estimation-selection problem and we cannot
conclude that the posterior estimates represent the best possible
model. This leads us to the following question: how can we perform model selection over all possible models efficiently?\\


In this paper, the main goal is to address these questions. To this end, (i) we propose a Bayesian approach for studying the
relationship among multiple brain regions by considering
point-mass priors, and (ii) we induce shrinkage on both
activation and connectivity parameters while capturing the high
frequency behavior of fMRI data. To take this particular behavior into account, we
propose a weakly informative default prior for the
variances of the state parameters that correspond to the
``activation" in the different brain regions. The prior induces
shrinkage and robustness suitable for high frequency fMRI data
with presence of sparsity, and produces
one-step-ahead proper posterior predictive results. The rest of the paper proceeds as follows. Section \ref{sec:priors} presents the formulation of
the proposed methodology. Section \ref{sec:sims} contains a simulation study
using multivariate dynamic models that illustrates the performance of our modeling approach, and in Section \ref{sec:apply} we apply the
proposed methodology to functional magnetic imaging data. Finally, a short discussion
is presented in Section \ref{sec:conclusions}.

\section{Modeling Approach}
\label{sec:priors}

Model selection has been one of the most active research areas in Bayesian analysis in recent years.
Mixture priors have been used in various settings as a variable selection-estimation tool in regression models (see for example \citeasnoun{George},
\citeasnoun{Clyde} and \citeasnoun{Raftery}). On the other hand, \citeasnoun{Huerta} use point-mass priors on the roots of the autoregressive polynomial
model to handle model uncertainty and unit roots in autoregressive models. \citeasnoun{bergardo} use point-mass priors for model selection to analyze DNA microarray
data. Among the most important and recent suggested approaches for model selection, we find the horseshoe prior by \citeasnoun{carvalho}, which arises from
considering a half-Cauchy distribution for the scale parameter of a Normal prior.
\citeasnoun{polson2} propose to use Inverted-Gamma densities for the scale parameter in a hierarchical fashion, and thus obtain a hypergeometric family
for modelling a dynamic autoregressive model.\\


In this work, we propose a Bayesian approach for studying the dynamic relationship between multiple brain regions. We describe the coupling or lead-lag
relationships between any pair of regions using point-mass mixture priors for the connectivity parameters as follows:
\begin{align}
\phi_{ij}\sim \pi N(0,\sigma^{2}_{ij})+ (1-\pi)\delta_{0}(\phi_{ij}),
\end{align}
such that the connectivity parameter is a non-zero drawn from the Normal prior with zero mean and variance $\sigma_{ij}^{2}=1/\tau_{ij}$ with
probability $\pi$, and zero with probability $1-\pi$. An advantage of this prior is that hypothesis testing and model selection can be performed at the same time. In contrast to the approach of \citeasnoun{ringo}, one important feature of the point-mass prior approach is that the assumption that connectivity parameters are equal to zero for some brain regions
is not necessary. The point-mass priors allow us to compute the posterior probability
of having a connectivity parameter equal to zero in a simple fashion. In other words, with the
point-mass approach we can not only obtain posterior inference on the connectivity parameters, but also consider all possible models for model comparison
purposes.

\subsection{Prior elicitation for the connectivity parameters} \label{sec:elicit}

In this section, we show the prior elicitation and corresponding simulation of the connectivity parameters. We consider this same elicitation in both the
simulation and application sections. We utilize prior information from results of the brain imaging data applications presented in \citeasnoun{ringo}, and use the proposal of \citeasnoun**{ideas} to elicit the connectivity parameters. Following \citename**{ideas},  we choose to find the $\text{Gamma}(c, d)$ prior for the precision
$\tau_{ij}=1/\sigma^{2}_{ij}$ by eliciting information about the first percentile of the sampling distribution. We assume a prior with mean zero and cumulative probability equal to 0.01 at -1 leading to $\tau_{0}\equiv (-1/\Phi^{-1}(0.01))^{-2} \approx 1.82$. By equating $\tau_{0}=(c-1)/d$ or equivalently $c=\tau_{0}d+1$, the prior for the precision parameter of the point-mass prior is 
$\tau_{ij} \sim \text{Gamma} (3.78,1.53)$.  \\

\begin{figure}[h]
\begin{center}
\includegraphics[scale=0.50
,keepaspectratio]{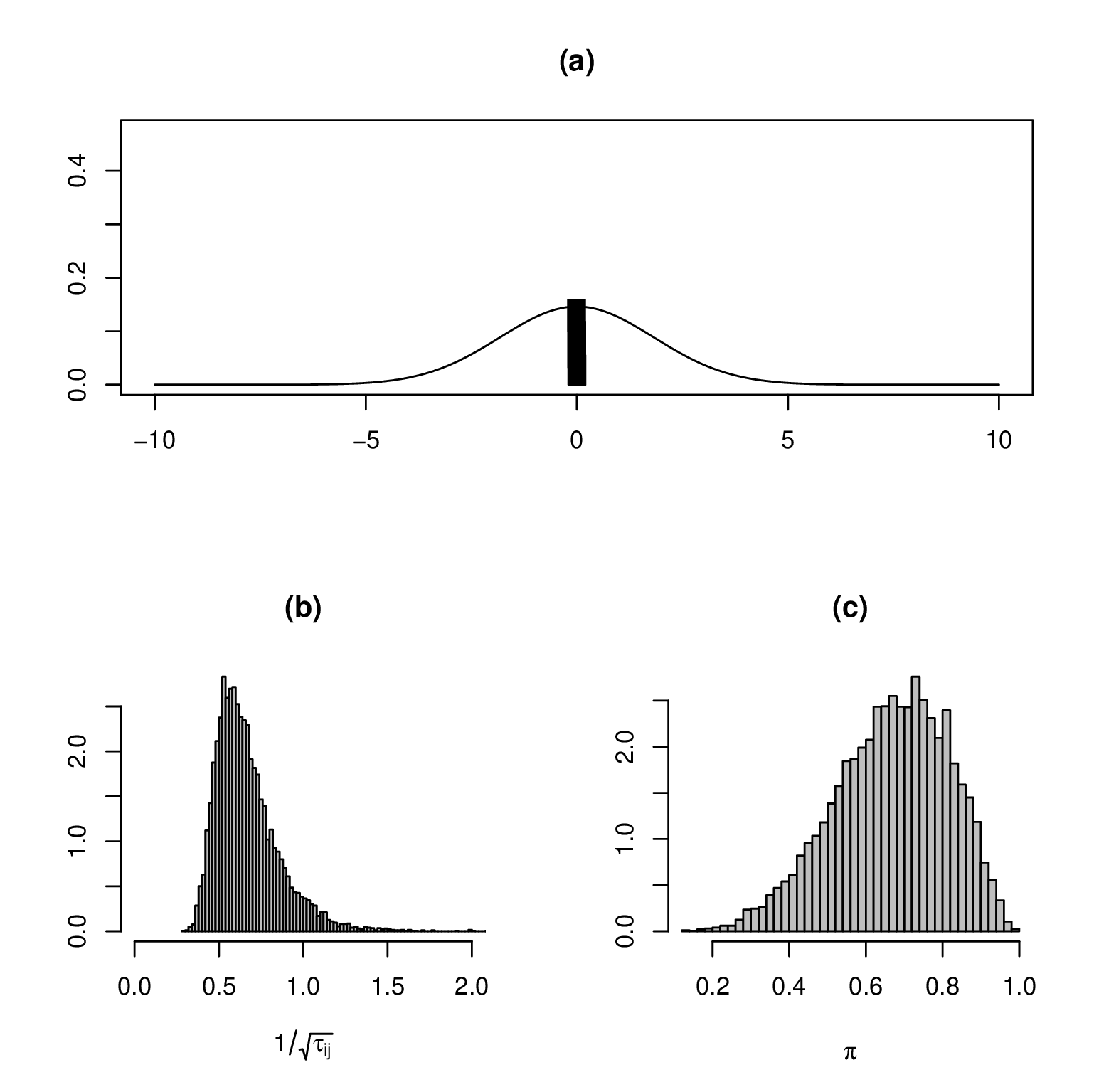}
\end{center}
\caption{(a) Point-mass prior density $(1-\pi)N(0,\tau_{ij}^{-1})$. The bar at zero corresponds to the mean of the weights $\pi$. (b) Prior density for the scale
$1/\sqrt{\tau_{ij}}$. (c) Prior for the weights $\pi$.}
\label{fig:priors}
\end{figure}

In order to specify the prior for the parameter $\pi$ of the point-mass prior, we use information from the results in \citename{ringo}. In their application, the number of 9 connectivity parameters different from zero is equal to 6. Therefore, we assume $\pi \sim \text{Beta}(a,b)$ with $a_{\pi}=6$ and $b_{\pi}=3$, so that the corresponding prior mean and standard deviation are $
E(\pi)=0.66$ and $\sqrt{V(\pi)}=0.149$, respectively. Figure \ref{fig:priors} displays the Normal prior for the point-mass prior, the corresponding variance and the weights using the elicitation described above.\\


\subsection{A weakly informative default prior for the state variances}

Weakly informative default prior choices for variances have been proposed in the past for Bayesian hierarchical models. For example, \citeasnoun{andrew} considers half-t prior distributions for scale parameters in hierarchical
models. The author proposes this weakly informative default prior to replace the very sensitive Inverse-Gamma$(\epsilon,\epsilon)$ ``non-informative'' conjugate prior in order to have
a limiting posterior distribution for hierarchical models.

We now present our proposal of a new weakly informative default
prior for the state variances in the general framework of Bayesian dynamic
linear models (BDLM).
The hierarchical definition of a BDLM for $t=1,\dots,T$ is,
\begin{align}\label{dlmu}
y_{t}|\boldsymbol{\theta}_{t}&\sim N(F_{t}\boldsymbol{\theta}_{t},V_{t}),\\ \notag \boldsymbol{\theta}_{t}|\boldsymbol{\theta}_{t-1}&\sim
N(G_{t}\boldsymbol{\theta} _{t-1},V_{t}W_{t}),\\ \notag \boldsymbol{\theta}_{t-1}|y_{1:t-1}&\sim N(m_{t-1},C_{t-1}),
\end{align}

\noindent where $\boldsymbol{\theta}_t$ corresponds to a vector of states of dimension $p$ varying smoothly over time and $F_{t}$ and $G_{t}$
are matrices of dimension $m \times p$ and $p \times p$, respectively. The parameter $V_{t}$ is the variance of the observation $y_t\mid \boldsymbol{\theta}_{t}$
and $V_tW_{t}$ is the variance of the state parameter $\boldsymbol{\theta}_{t}\mid\boldsymbol{\theta}_{t-1}$. In turn, $m_t$ and $C_t$ correspond
to the posterior mean and posterior variance of the state parameter $\theta_t$ given $y_{1:t-1}$. For simplicity, we let $y_{t}$ be the value of an univariate
time series at time $t$ with $\theta_{t}$ corresponding to an unobservable state vector. Also, we consider $F_{t}=1$, $G_{t}=G=\phi $, $V_{t}=V=\sigma^{2}$ and
$W_{t}=\tau^{2}_{t}$. The model (\ref{dlmu}) is studied in the seminal book of \citeasnoun{westt}, where it is assumed that the state variance $W_{t}$ is unknown
and discount factors are proposed for modelling it. \\

Let us consider the one-step-ahead predictive distribution of $y_{t}$ given $y_{1:t-1}$ for the model
in (\ref{dlmu}), which follows a Gaussian distribution with mean and variance given by
\begin{align} \notag
f_{t}&=m_{t-1},\\ \notag Q_{t}&=C_{t-1}+\sigma^{2}+\tau^{2}_{t}\sigma^{2}. \notag
\end{align}
Assume $\sigma^{2(*)}=C_{t-1}+\sigma^{2}$ and $\sigma^{2}=1$ for
simplicity. Then  the density function of the
one-step-ahead predictive distribution is as follows:

\begin{align}\label{form}
p(y_{t}|y_{1:t-1},\sigma^{2(*)},\lambda_{\theta}^{-1})\propto
\dfrac{1}{\sqrt{\sigma^{2(*)}+\lambda_{\theta}^{-1}}}\exp\left\{-\frac{1}{2}\dfrac{(y_{t}-m_{t-1})^{2}}{\sigma^{2(*)}+\tau^{2}_{t}}\right\}.
\end{align}
where $\lambda_{\theta}=1/(\tau^{2}_{t}\sigma^{2})$ is the state
precision. The Jeffreys prior
$p(\sigma^{2(*)})\propto\sigma^{-2(*)}$ poses no issues. However,
analogously to \citeasnoun{andrew}, in the hierarchial model case
if we consider the Jeffreys prior $p(\tau^{2}_{t})\propto
\tau^{-2}_{t}$, we have that the density function in (\ref{form})
is positive at $\tau^{2}_{t}=0$ and therefore $p(\tau^{2}_{t})$
fails to be integrable at the origin. Also, the
conjugate Inverse-Gamma$(\epsilon,\epsilon)$
prior is very sensitive to choices of very small values of
$\epsilon$ leading to an improper posterior one-step-ahead
predictive density. \\

On the other hand, the Beta prime density has
been considered by different authors as a default prior for
variances in Bayesian model selection (see \citeasnoun{steel} and
\citeasnoun**{gprior}), hierarchical models \cite{polson},
and for modelling outliers and structural breaks in BDLMs
\cite{fuquenep}. The Beta prime density with shape
parameters $p$ and $q$ and scale dynamic parameter $\beta_{t}$ is
described as,
\begin{equation}
\pi(\tau^{2}_{t})=\frac{\Gamma(p+q)}{\Gamma(p)\Gamma(q)}\frac{1}{\beta_{t}}\frac{\left(\dfrac{\tau^{2}_{t}}{\beta_{t}}\right)^{p-1}}{\left(1+\dfrac{\tau^{2}_{t}}{\beta_{t}}\right)^{p+q}},
\;\;\;\; \tau>0,
\end{equation}
where $\Gamma(\cdot)$ corresponds to the gamma function. Here, for mathematical properties and computational simplicity, we propose the use of a Beta prime density
with $p=1$ and $q=(\upsilon_{t}-1)/2$:
\begin{equation}\label{scbeta2}
p(\tau^{2}_{t})\propto\left(1+\dfrac{\tau^{2}_{t}}{\beta_{t}}\right)^{-(\upsilon_{t}+1)/2}.
\end{equation}
Combining the density (\ref{form}) and the prior (\ref{scbeta2}), we have that $p(y_{t}|y_{1:t-1},\sigma^{2(*)},\lambda_{\theta}^{-1})$ is defined when
$\tau^{2}_{t}\rightarrow 0$. For the case
 $\tau^{2}_{t}\rightarrow \infty$, the exponential term in
(\ref{form}) is less than or equal to 1. For the remaining term, we have that
$(1+\tau^{2}_{t}/\sigma^{2(*)})^{-1/2}(1+\tau^{2}_{t}/\beta_{t})^{-(\upsilon_{t}+1)/2}$ is integrable and hence
$p(y_{t}|y_{1:t-1},\sigma^{2(*)},\lambda_{\theta}^{-1})$ is proper. \\

The Beta prime distributions considered here induce one-step-ahead proper posterior
predictive results and sampling from these priors is straightforward due to the mixing Gamma property
$\tau^{2}_{t}\sim
\text{Gamma}(1,\beta_{t}/\rho_{t})$ and $\rho_{t}\sim \text{Gamma}((\upsilon_{t}-1)/2,1)$.
Also, by definition, the Beta prime for the scale parameter $\lambda_{\theta}=\tau^{-2}_{t}$ has shape parameters $p=(\upsilon_{t}-1)/2$ and $q=1$
and a dynamic scale parameter $1/\beta_{t}$. The priors for the observation and state variances are summarized in the display below.
To make the inference procedure feasible, we use Monte Carlo Markov Chain (MCMC) methods. The summary of the algorithm is available in Appendix A of the supplementary materials. 

\begin{align}
V_{t}^{-1}&=1/\sigma^{2}=\lambda_{y}, \quad p(\sigma^{2})\propto 1/\sigma^{2},\\ \notag W_{t,i}^{-1}&=\lambda_{y}\lambda_{\theta,i}\omega_{\theta,t_{i}}, \quad  i=1,\dots,p \\ \notag
\omega_{\theta,t_{i}}|\nu_{\theta,t_{i}}&\sim \text{Gamma}(\nu_{t_{i}}/2,\nu_{t_{i}}/2),\\ \notag \lambda_{\theta,i}&\sim
\text{Gamma}((\nu_{t_{i}}-1)/2,\rho_{t_{i}}/\beta_{t_{i}}),\\\notag \rho_{t_{i}}&\sim \text{Gamma}(1,1),\\ \notag \beta_{t_{i}}&\sim \text{Gamma}(1,\xi_{t,i}),\\
\notag \xi_{t,i}&\sim \text{Gamma}(1,1),\\ \notag \upsilon_{\theta,t_{i}}&\sim \text{Multinomial}(1,\varphi_{i}),\\ \notag \varphi_{i}&\sim
\text{Dirichlet}(\alpha_{i}),
\end{align}
Under this formulation, the state variances follow a Student's t-distribution with $\nu_{t}$ degrees of freedom by assuming $\tau_{t}^{2}|\lambda_{\theta,i},\omega_{\theta,t_{i}}\sim N(0,\sigma^{2}(\lambda_{\theta,i}\omega_{\theta,t_{i}})^{-1})$,
where the degrees of freedom follow a multinomial distribution  as assumed by
\citeasnoun{petris}. The marginal prior for the states can be found in a closed form as follows: (see proof in Appendix B - supplementary material
)

\begin{defin}

The  marginal prior of the states in model \ref{dlmu} using the variance formulation in \ref{scbeta2} is as follows:
\begin{align} \label{eq:priorstate}
\pi(\theta_{t}|G_{t}\theta_{t-1},\sigma,\nu_{t},\beta_{t})&=\dfrac{\nu_{t}-1}{2\sqrt{\sigma\nu_{t}\beta_{t}}\left(1+\dfrac{|\theta_{t}-G_{t}\theta_{t-1}|}{\sqrt{\sigma\nu_{t}\beta_{t}}}\right)^{\nu_{t}}}.
\end{align}

\end{defin}

\begin{figure}[h]
\begin{center} 
\label{fig:stateprior}
\includegraphics[scale=0.8,keepaspectratio]{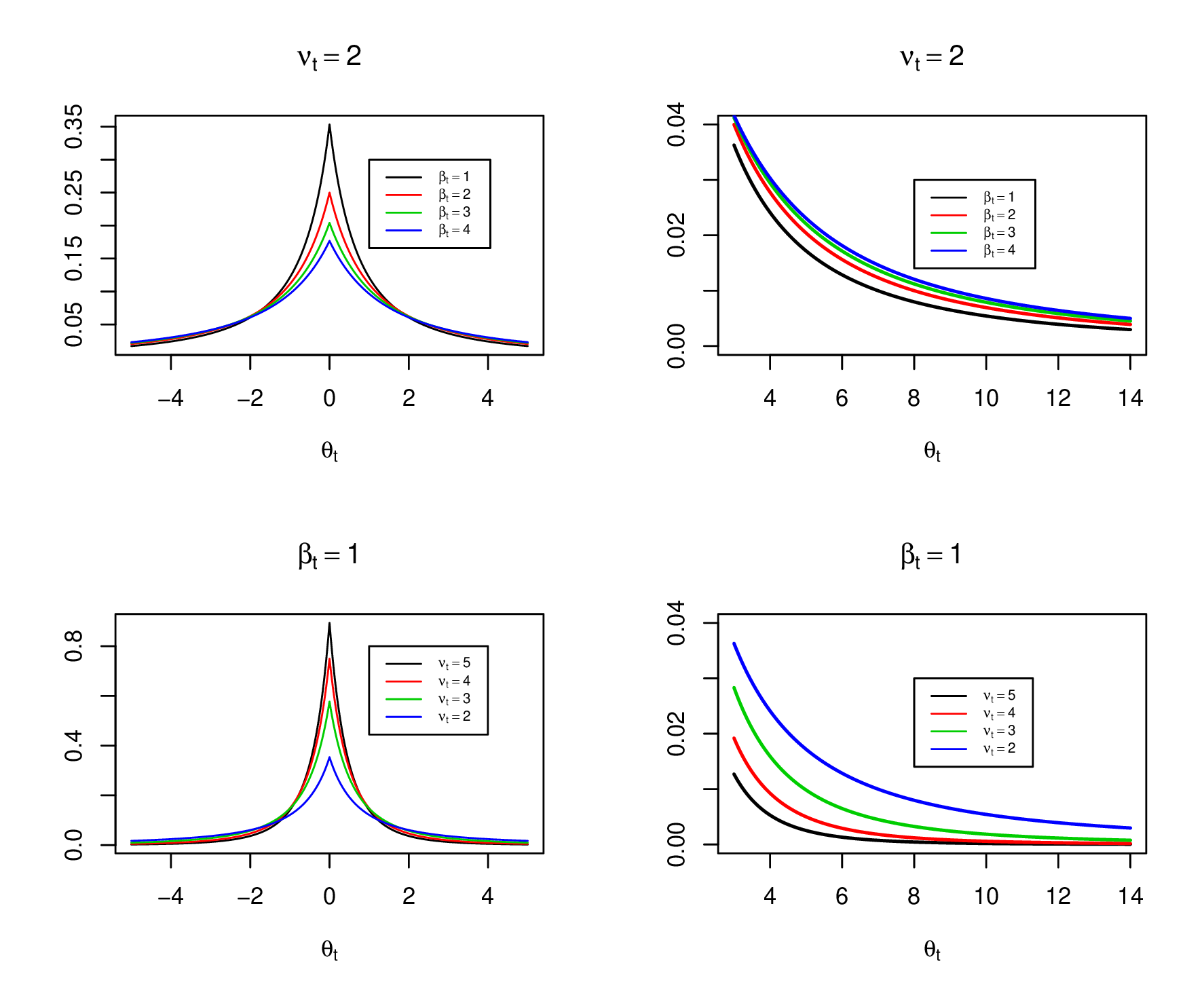} 
\caption{Comparison of marginal priors for the states considering different values of the hyperparameters $\nu_{t}$ and $\beta_{t}$.}
\end{center}
\end{figure}

Particular cases of priors as the one in equation (\ref{eq:priorstate}) have appeared repeatedly in the literature over the years under various names (Linnik, Meridian,
double-Pareto, generalized t and normal- gamma), e.g. \citeasnoun{Devroye}, \citeasnoun{Armagan}, \citeasnoun{Kawata}, \citeasnoun**{Lee}  and \citeasnoun{Griffin}.
Also, the particular case when $\sigma=\nu_{t}=G_{t}=1$, corresponds to the Scaled-Beta-Cauchy prior proposed by \citeasnoun{fuquenep}. The $g$-prior used in
\citeasnoun{steel} seems to be in the same class, except that the prior for $\tau^{2}_{t}$ is improper.  Figure \ref{fig:stateprior} illustrates how the density is more
heavy-tailed when the degrees of freedom $\nu_{t}$ increases, the marginal prior becomes weakly informative and the variance increases with $\beta_{t}$. Moreover,
to avoid over-shrinking of the states and to learn fully automatically, we also introduce priors for the
parameters in equation (\ref{scbeta2}). \\

Note that shrinkage is also induced for the
connectivity parameters $\phi$, where a marginal prior with a similar form to the one in  (\ref{eq:priorstate}) could be obtained by using the full conditional distribution of $\phi$
and integrating out the state variances. Also, when $\nu_{t} \rightarrow \infty$, the prior becomes more similar to a Normal prior in the first level
of the hierarchical model, although with a Student's t tail behavior. Therefore, the novelty of our approach is not only proposing a default state variance prior
suitable for detecting sparse state-signals of BDLMs applied to fMRI data. We also induce shrinkage in the estimation of the autoregressive coefficient
parameter.

\begin{figure}[h]
\begin{center}
\includegraphics[scale=0.5,keepaspectratio]{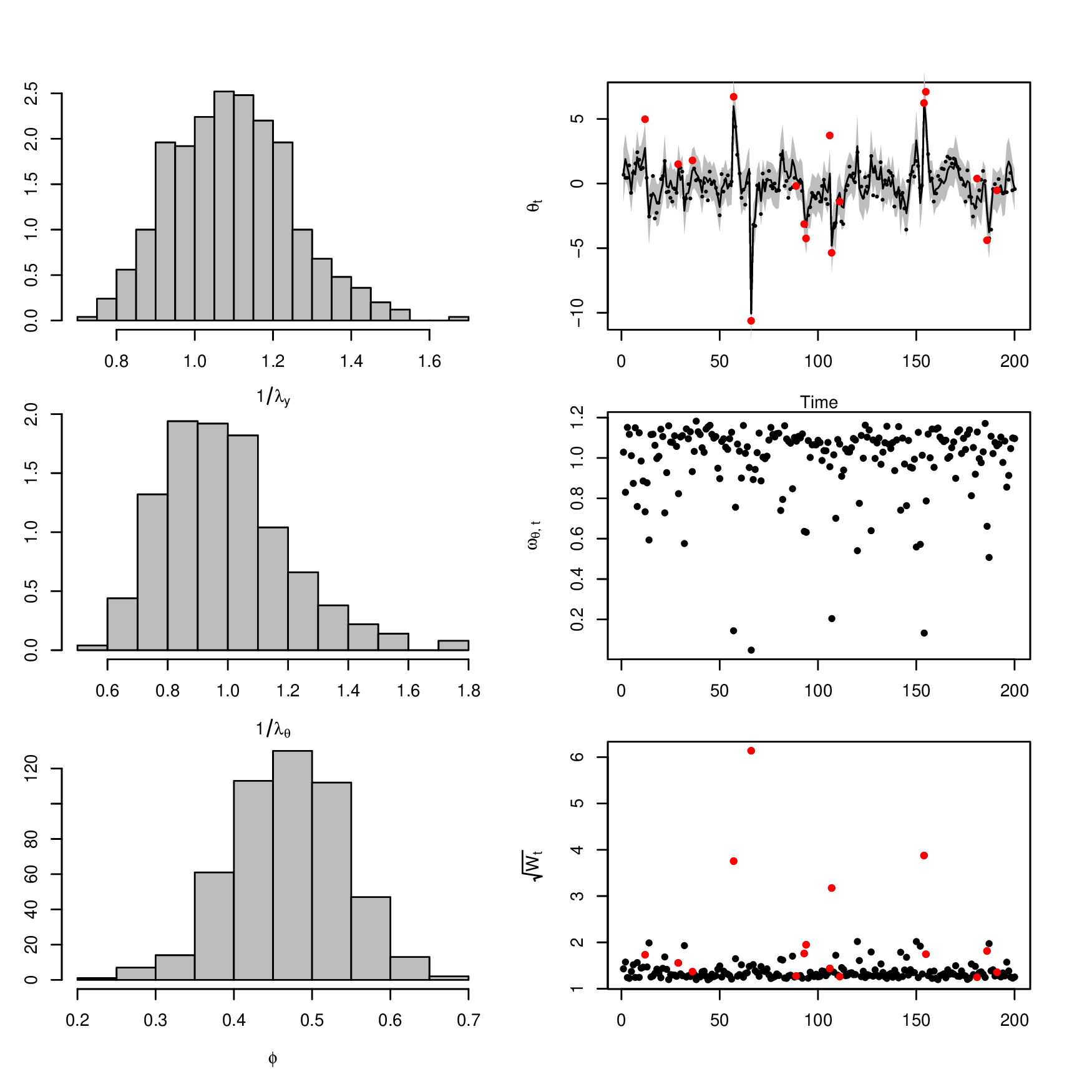}
\caption{Left part: posterior mean densities of $V=1/\lambda_{y}$, $1/\lambda_{\theta,i}$ and $\phi$. Right part: posterior means
of $(\theta_{t}|y_{t})$ over time with their corresponding credible bands (hatched area), the posterior mean of the state variances
$W_{t}=1/(\lambda_{y}\lambda_{y}\omega_{\theta,t})$ and the posterior mean of $\omega_{\theta,t}$. Signal/noise=1.} 
\label{fig:simula5}
\end{center}
\end{figure}

\begin{figure}[h]
\begin{center} 
\includegraphics[scale=0.48,keepaspectratio]{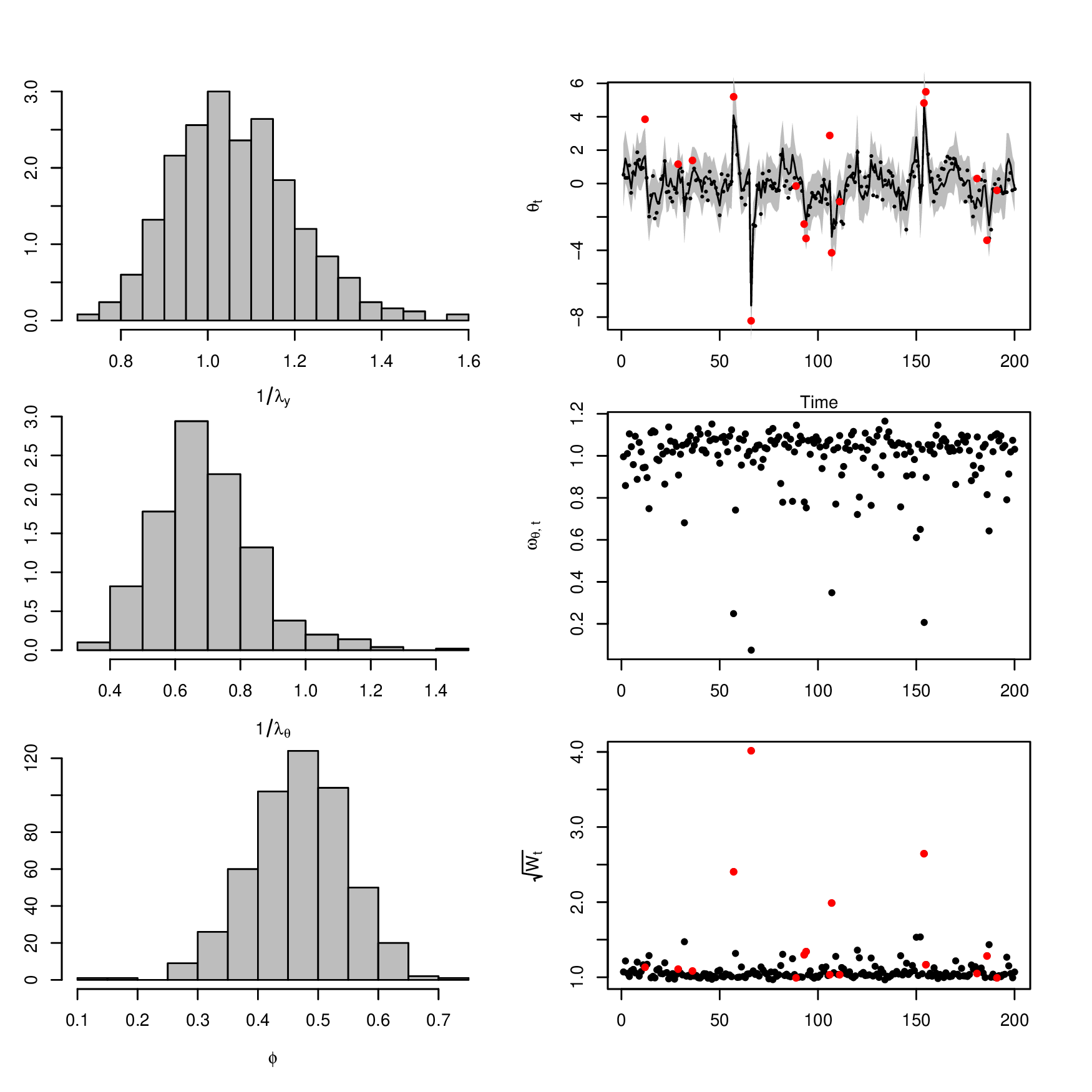} 
\caption{Left part: posterior mean densities of $V=1/\lambda_{y}$, $1/\lambda_{\theta,i}$ and $\phi$. Right part: posterior means
of $(\theta_{t}|y_{t})$ over time with their corresponding credible bands (hatched area), the posterior mean of the state variances
$W_{t}=1/(\lambda_{y}\lambda_{y}\omega_{\theta,t})$ and the posterior mean of $\omega_{\theta,t}$. Signal/noise=0.6.} 
\label{fig:simula3}
\end{center}

\begin{center} 
\includegraphics[scale=0.53, keepaspectratio]{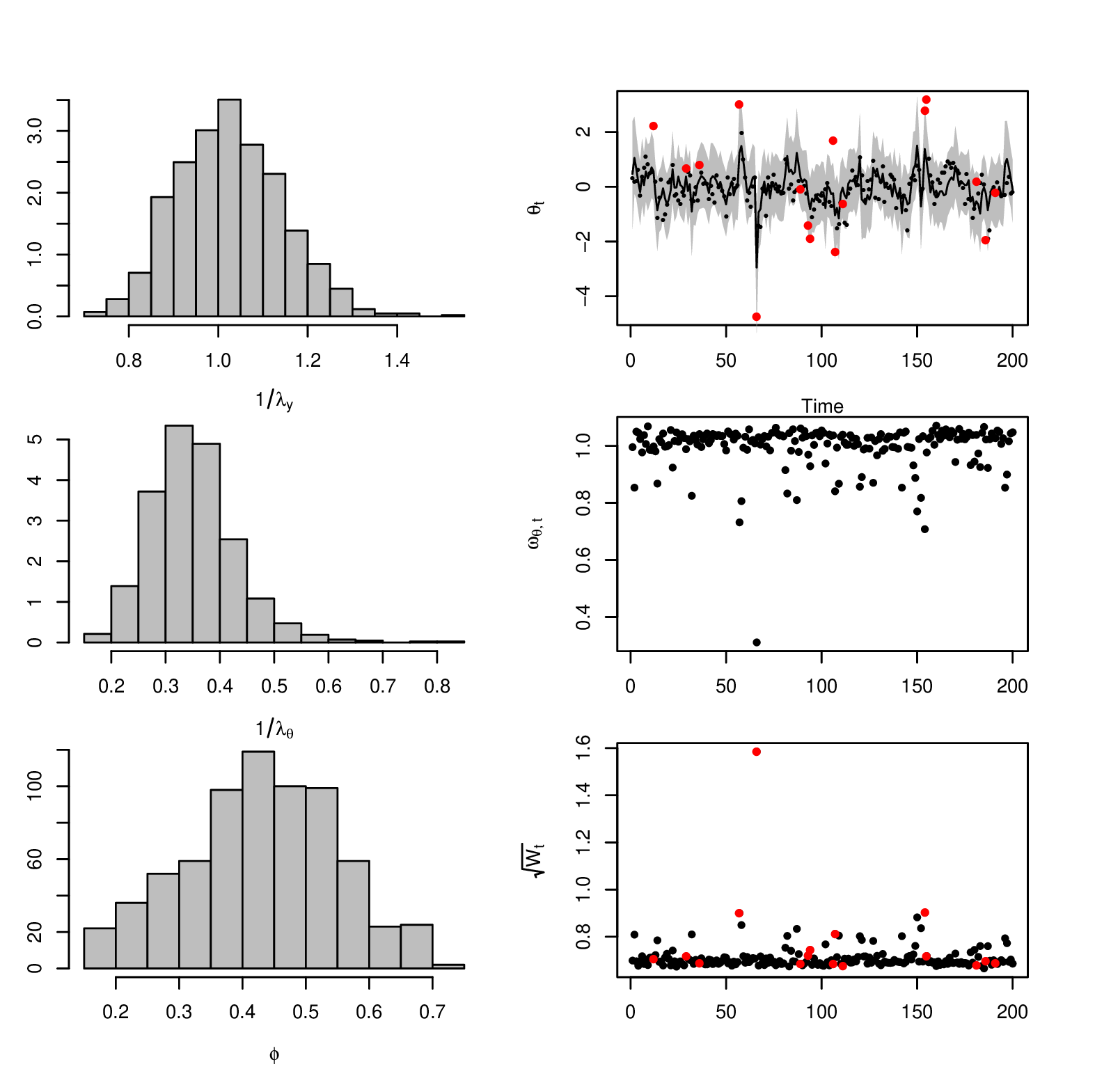} 
\caption{Left part: posterior mean densities of $V=1/\lambda_{y}$, $1/\lambda_{\theta,i}$ and $\phi$. Right part: posterior means
of $(\theta_{t}|y_{t})$ over time with their corresponding credible bands (hatched area), the posterior mean of the state variances
$W_{t}=1/(\lambda_{y}\lambda_{y}\omega_{\theta,t})$ and the posterior mean of $\omega_{\theta,t}$. Signal/noise=0.2.} 
\label{fig:simula1}
\end{center}
\end{figure}

We present now synthetic examples to illustrate the performance of our proposed weakly informative prior. We consider the following BDLM:
\begin{align}
y_{t}&=\theta_{t}+v_{t},  & \theta_{t}&=\phi\theta_{t-1}+w_{t},
\end{align}
where the sparse signals $w_{t}$, $t=1,\dots,T$, follow a two component Normal mixture model given by
\begin{equation}
w_{t}\sim \pi N(0,V) + (1-\pi)N(0,\kappa W),
\end{equation}
and $v_{t}\sim N(0,V)$. We consider $\kappa=20$, $\phi=0.5$, $W/V=\{1, 0.6, 0.2\}$ and $\pi=0.9$. The Markov Chain Monte Carlo scheme, where we
also use the Forward Filtering Backward Sampling (FFBS) algorithm proposed in \citeasnoun{Fruwirth} for posterior inference purposes is presented in Appendix A - supplementary material.
We reached convergence of all parameters in 5,000 iterations after a burn-in period of 2,000 iterations with a thinning period of 10. We spent
approximately 50 minutes to obtain the results using the \citeasnoun{RR} program and a PC with Intel(R) Xeon(R) 2.80 GHZ and 4 GB RAM. Figures \ref{fig:simula5} to \ref{fig:simula1}
illustrate the results. In the right panels, the red circles correspond to values from the $N(0,\kappa V)$ mixture component and the black circles correspond to
values from the $N(0,W)$ component. We can see in all cases that the posterior distributions of $1/\lambda_{y}$ and $1/\lambda_{\theta,i}$ reproduce the true
parameters. The posterior mean density of $\phi$ represents nicely the true value and the corresponding probability in all cases. The posterior mean of the state
variances $W_{t}=1/(\lambda_{y}\lambda_{\theta,i}\omega_{\theta,t})$ and the posterior mean of the latent parameters $\omega_{\theta,t}$ properly identify the sparse state signals with values $\omega_{\theta,t}< 1$. The Figures illustrate how shrinkage is induced under the small values of the $\phi$ parameter.

\section{Simulation study} \label{sec:sims}

We explore two different modeling settings on simulated data. We first fitted
a model where the state precisions, $\lambda_{\theta,i}$, are
fixed. In the second model we consider the precisions
$\lambda_{\theta,i}$ unknown and we use the proposed weakly
informative prior for the state variances presented
in the last section. For both settings, we consider three time series of
size $T=285$ and we use different values of the signal/noise
ratio $\lambda_{\theta,i}^{-1}/\lambda_{y,i}^{-1}=\left\{0.5,1,2\right\}$
and $\lambda_{\theta,i}^{-1}=1$ in order to study
the performance of a model with sparse state parameters. The model and parameter values used in the simulation (also applied in the last section) are the following follows:

\begin{align*} \small
\notag  \boldsymbol{y}_{t}=\begin{pmatrix}
  y_{t,1}\\
  y_{t,2}\\
  y_{t,3}
\end{pmatrix},
 \quad \boldsymbol{F}_{t}=\begin{pmatrix}
1 & 0 & 0 &  x_{t,1} & 0 & 0\\ 0 & 1 & 0  & 0 & x_{t,1} & 0\\ 0 & 0 & 1 &  0 & 0 & x_{t,1}
\end{pmatrix}, \\ \\
 \boldsymbol{G}_{t}=
\begin{pmatrix}\notag
1 & 0 & 0 & 0 & 0 &  0 \\ 0 & 1 &  0 & 0 &
  0 & 0\\
0 &
  0 & 1 & 0 & 0 &  0 \\
0 & 0 &  0  & \phi_{11}x_{t-1,1} & \phi_{12}x_{t-1,2} & \phi_{13}x_{t-1,3} \\ 0 & 0 &  0 & \phi_{21}x_{t-1,1} & \phi_{22}x_{t-1,2} & \phi_{23}x_{t-1,3} \\ 0 & 0 &
0 &  \phi_{31}x_{t-1,1} & \phi_{32}x_{t-1,2} & \phi_{33}x_{t-1,3} \\
\end{pmatrix},
\end{align*}

\begin{align*}\notag\boldsymbol{V}_{t}&=\boldsymbol{V}=diag\begin{pmatrix}
 \lambda_{y,1}^{-1}, & \lambda_{y,2}^{-1}, &,
\lambda_{y,3}^{-1}\notag
\end{pmatrix},
\\\notag
 \boldsymbol{W_{t}}&=\boldsymbol{W}=diag\begin{pmatrix} \notag
0, 0, 0,  (\lambda_{y,1}\lambda_{\theta,1})^{-1}, & (\lambda_{y,2}\lambda_{\theta,2})^{-1}, &, (\lambda_{y,3}\lambda_{\theta,3})^{-1}\notag
\end{pmatrix},
\\\notag
\boldsymbol{\theta}_{t}^{'}&=\begin{pmatrix}
  \alpha_{1}, &
  \alpha_{2}, &
  \alpha_{3}, &
  \theta_{t,1}, & \theta_{t,2}, & \theta_{t,3}
\end{pmatrix}.
\end{align*} \\

Table \ref{tab:phis} displays the values of the connectivity parameters used to simulate the data. These values are based on the results of \citeasnoun{ringo} where some connectivity regions are close to zero.
\begin{table}[htb]
\begin{center}
\caption{True values for the connectivity regions in the simulation study.} \small{ \label{simula2}
\label{tab:phis}
\renewcommand{\arraystretch}{1.5}
\begin{tabular}{cccccccccc}\\
 \hline
 $\phi_{11}$ & $\phi_{12}$ & $\phi_{13}$ & $\phi_{21}$ & $\phi_{22}$ & $\phi_{23}$ &  $\phi_{31}$ & $\phi_{32}$ & $\phi_{33}$\\\hline\hline
0 & -0.1495 & -3.0382 & 0 & -0.8365 & -0.2667 &  0.4179 & 0.1365 & 0\\
\end{tabular}
}
\end{center}
\end{table}
We use a non-informative Gamma prior for the observational precisions with hyperparameters $a_{y,i}=0.001$ and $b_{y,i}=0.001$, and a
Beta$(a_{\pi},b_{\pi})$ prior for the weights $\pi$ with hyperparameters $a_{\pi}=6$ and $b_{\pi}=3$. We assume the weakly informative default prior
for the state precisions. For the connectivity parameters $\phi_{ij}$, we consider the point-mass prior with the elicitation presented in Section \ref{sec:priors}. Using standard methods such as the autocorrelation function, time series traces and cumulative estimates of the quantiles, we verified the convergence of all
parameters using a burn-in
period of 10000 iterations with 30000 subsequent iterations to generate
the estimated posterior distributions (see MCMC scheme in Appendix A - supplementary material). To have a measure of the forecasting accuracy, we use two common criteria called the
mean absolute deviation (MAD) and the mean square error (MSE), which are defined as
\begin{align*}
MAD&=\frac{1}{285}\sum_{i=1}^{3}\sum_{t=1}^{T}|e_{i,t}| & MSE&=\frac{1}{285}\sum_{i=1}^{3}\sum_{t=1}^{T}e_{i,t}^{2},
\end{align*}
where $e_{i,t}=y_{i,t}-(\alpha_{i}^{s}+F^{'}_{t}\theta^{s}_{i,t})$,
for $\alpha_{i}^{s}$ and $\theta^{s}_{i,t}$, the simulated parameters.

\clearpage

\begin{table}[htb]

\begin{center}
{ \caption{Forecasting accuracy measures for two models settings where $\lambda_{\theta,i}$
is considered known or unknown. The data was simulated using $\lambda_{\theta,i}=1$.}\label{tab:simula2}
\renewcommand{\arraystretch}{2.5}
\begin{tabular}{ccc}
Signal/noise ratio & MAD & MSE
\\\hline
$\lambda_{\theta,i}^{-1}/\lambda_{y,i}^{-1}=0.5$ ; $\lambda_{\theta,i}$ unknown &   4.370 & 3.458\\
 \hline
$\lambda_{\theta,i}^{-1}/\lambda_{y,i}^{-1}=0.5$ ; $\lambda_{\theta,i}$ known & 4.010 & 3.184\\
 \hline
$\lambda_{\theta,i}^{-1}/\lambda_{y,i}^{-1}=1$ ; $\lambda_{\theta,i}$ unknown &   2.403 & 1.904
\\
 \hline
$\lambda_{\theta,i}^{-1}/\lambda_{y,i}^{-1}=1$ ; $\lambda_{\theta,i}$ known & 2.752 & 2.185
 \\
 \hline
$\lambda_{\theta,i}^{-1}/\lambda_{y,i}^{-1}=2$ ; $\lambda_{\theta,i}$ unknown & 1.736 & 1.374\\
 \hline
$\lambda_{\theta,i}^{-1}/\lambda_{y,i}^{-1}= 2$ ; $\lambda_{\theta,i}$ known &   1.812 & 1.431\\
 \hline
\end{tabular}
}
\end{center}
\end{table}
Table \ref{tab:simula2} shows the results of the measures of accuracy in the simulation. We are interested in comparing MAD and MSE for the same model when the state precisions are known or unknown in order to evaluate the performance of the proposed weakly informative prior. According to the results, using the proposed weakly informative prior for the state
precisions could be a good choice given that the MAD and MSE values are similar to those obtained when the precisions are known for the different signal/noise ratios. In Appendix C - supplementary material, in Figures 1, 5, 9, 13, 17 and 21, we can see how most of the values of the posterior means for the connectivity parameters are close to the true values. We can also see in Figures 2, 3, 6, 7, 10, 11, 14, 15, 18, 19, 22 and 23 that the posterior
densities of trends, state and observational variances are concentrated around the true values. Similarly, according to Figures 4, 8, 12, 16, 20 and 24 the true
state parameters are generally within the 95\% simulated credible intervals.

\section{Application: fMRI data}  \label{sec:apply}

This section presents the application of the proposed methodology for researching the mechanism of attentional control with fMRI time series from a single subject.
We consider the same example shown in \citeasnoun{ombao} and \citeasnoun{ringo}, who consider state-space models for studying the dynamic relationship between
multiple brain regions. According to \citeasnoun{Banish}, three systems involve attentional control: (1) the task-relevant process system, which involves the
task-relevant stimulus dimension; (2) the task-irrelevant processing system, which allows to process the task-irrelevant stimulus dimension; and (3) a source of
control that develops the top-down selection bias, which may increase the neural activity within the task-relevant processing system and/or may suppress the
neural activity within the task irrelevant processing system. Many applications have found the dorsal prefrontal cortex to be a main source of the attention
control.

\subsection{Experimental design}

\textbf{Data acquisition.}  A GE Signa magnetic resonance imaging system equipped for echoplanar imaging (EPI) was used for data acquisition (see
\citeasnoun{milann}). Eleven right-handed native English-speaking participants (7 men and 4 women, ranging in age from 18 to 30) were included in the study. For
each run, a total of 300 EPI images were acquired (TR = 1517 ms, TE = 40 ms, flip angle $90^{\circ}$), each consisting of 15 contiguous slices (thickness 7 mm,
in-plane resolution 3.75 mm), parallel to the AC-PC line. A high-resolution 3D anatomical set (T1-weighted three-dimensional spoiled-gradient echo images) was
collected for each participant, as well as T1 weighted images of our functional acquisition slices. The head coil was fitted with a bite bar to minimize head
motion during the session. Stimuli were presented on a goggle system designed by Magnetic Resonance Technologies. In the experiment, two phases were explored:

\begin{itemize}
\item \textbf{Learning phase}. The subject learned to associate each of three unfamiliar shapes with one of three color words (i.e. ``BLUE", ``YELLOW" or
    ``GREEN") and at the end of this phase it was verified that participants could correctly provide the name of the three shapes with 100\% accuracy. Next,
    the shapes were presented in white without their associated words, one at the time in random order. Finally, the participants were instructed to practice
    naming each shape subvocally with its corresponding word. Each shape was presented a total of 32 times.

\item \textbf{Test phase}. In this phase, blue, yellow
  and green ink colors were used and two types of trials were presented:
\enlargethispage{1cm}

\begin{itemize}
\item \textbf{The interference trial}. In the interference trial the shape was printed in an ink color incongruent with the color used to name the shape.
    \item \textbf{The neutral trial}. In the neutral trial the shape was printed in white, which was not a color name for any of the shapes.
\end{itemize}

\end{itemize}

A block design was used where the block of neutral trials was alternated with the block of interference trials. We have 6 blocks of neutral and interference
trials, where each block consists of 18 trials presented at a rate of one trial each 2 seconds. Each trial consisted of a 300 milliseconds fixation cross by a
1,200 millisecond presentation of the stimulus (shape) and a 500 millisecond inter-trial interval. Finally, participants were instructed to subvocally name each
shape with the corresponding color from the learning phase ignoring the ink color in which the shape was presented. Subvocalization (characterized by the
occurrence in the mind of words in speech order with or without inaudible articulation of the speech organs) was utilized in an effort to avoid possible motion
artifacts. Figure \ref{fig:stimulus} displays the stimulus and hemodynamic response function of this experiment.

\subsection{The three regions of interest}

\begin{figure}[h]
\begin{center}
\includegraphics[scale=0.7
,keepaspectratio]{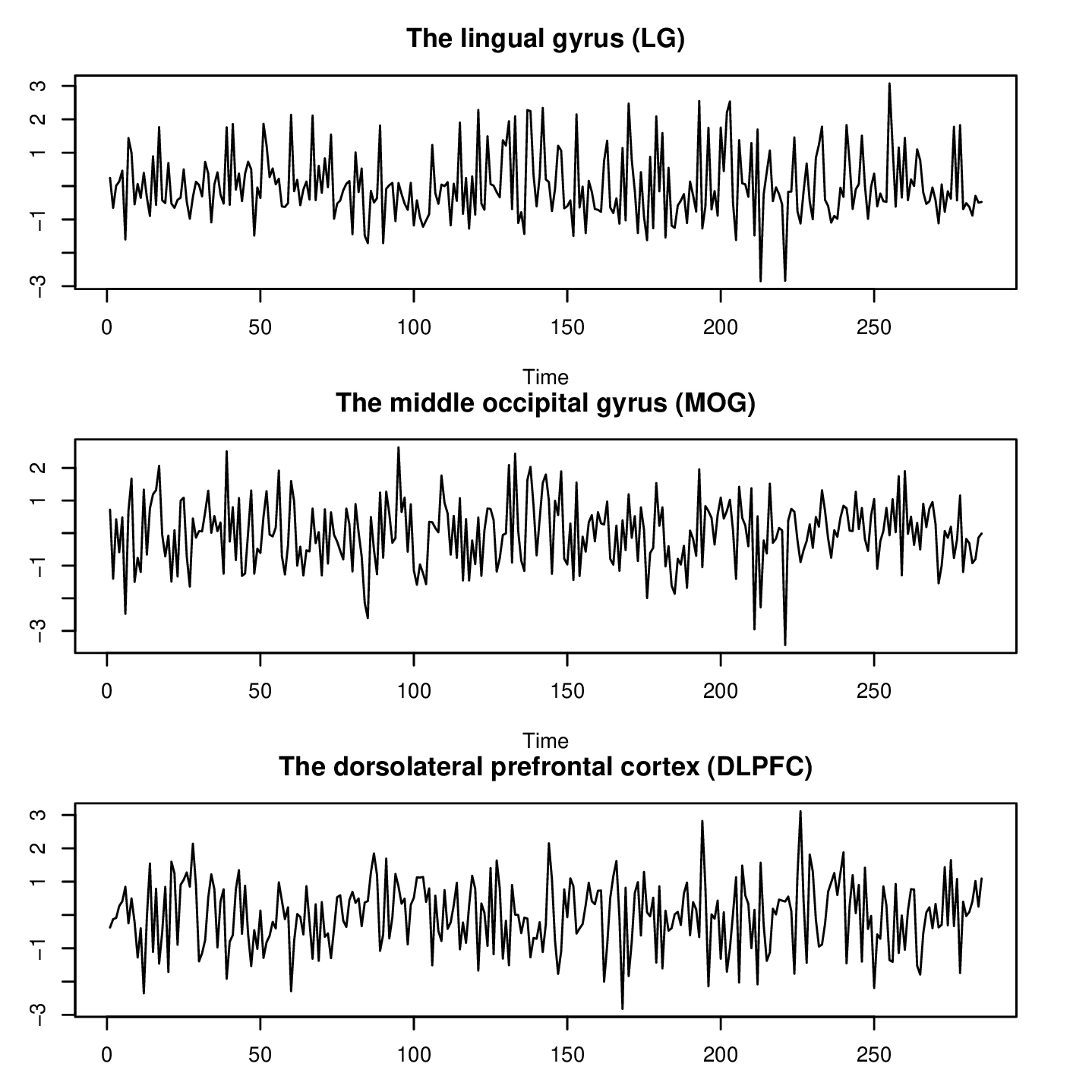} \caption{fMRI time series data for the application.}
\label{fig:regions}
\end{center}
\end{figure}

We are interested in the attention control network that reflects the brain's ability to discriminate between relevant and irrelevant information in tasks that require a certain level of concentration. The lingual gyrus, the middle occipital gyrus and the dorsolateral prefrontal cortex were selected. The
lingual gyrus (LG) is a visual area sensitive to color information which can be used as a site for processing task-irrelevant information  (i.e., the ink color
\cite**{kelley}). The middle occipital gyrus (MOG) is also a visual area sensitive to shape information and it represents a site for processing task-relevant
information (i.e., the shapes form). The dorsolateral prefrontal cortex (DLPFC) is selected to represent the source of attentional control. Figure \ref{fig:regions} displays the
standardized time series of the three regions of interest. The three time series regions were detrended using a linear smoother which is roughly a linear
regression fitted to the $k$-nearest neighbors of a given point and it is used to predict the response at that point.

We consider the same multivariate dynamic model presented in Section \ref{sec:sims} where the three regions are the lingual gyrus (LG), the middle occipital gyrus (MOG), and the dorsolateral prefrontal cortex (DLPFC), respectively. For instance, $\phi_{11}$ represents the self-feedback in the LG region, and $\phi_{12}$ characterizes the coupling relationship between the LG and MOG regions. In the MCMC algorithm, we obtained convergence of all parameters using 30000 iterations after a
burn-in period of 10000 iterations and a thin of 4 where different initial values were considered. We used a non-informative Gamma prior for the observational
precisions with hyperparameters $a_{y,i}=0.001$ and $b_{y,i}=0.001$. The state variances are modeled using the proposed weakly informative prior. For the connectivity
parameters, we considered the point-mass prior with the elicitation presented in section \ref{sec:elicit} for the precision and the weights. 

\begin{table}[htb]
\caption{Posterior Mean, posterior standard deviation and posterior probability of $\phi_{ij}=0$.}\label{simula2}
\renewcommand{\arraystretch}{1.1}
\begin{center}
\begin{tabular}{cccccccc}\\
\hline\hline Parameter & Posterior mean & Posterior SD & $P(\phi_{ij}=0|\text{data})$\\
 \hline\hline
 $\phi_{11}$ & 0 & 0 & 1.00\\
$\phi_{12}$ & -0.0335 & 0.09 & 0.99\\ $\phi_{13}$ & -5.4126 & 0.64 & 0.00\\ $\phi_{21}$ & 0 & 0 & 1.00\\ $\phi_{22}$ & 0.0308 & 0.01 & 0.99\\ $\phi_{23}$ & -4.940
& 0.71 & 0.00\\ $\phi_{31}$ & -0.091 & 0.05 & 0.99\\ $\phi_{32}$ & -0.1250 & 0.06 & 0.98\\ $\phi_{33}$ & 0.3221 & 0.18 & 0.61\\\hline\hline
\end{tabular}
\end{center}
\end{table}

\begin{figure}
\begin{center}
\includegraphics[scale=0.5,keepaspectratio]{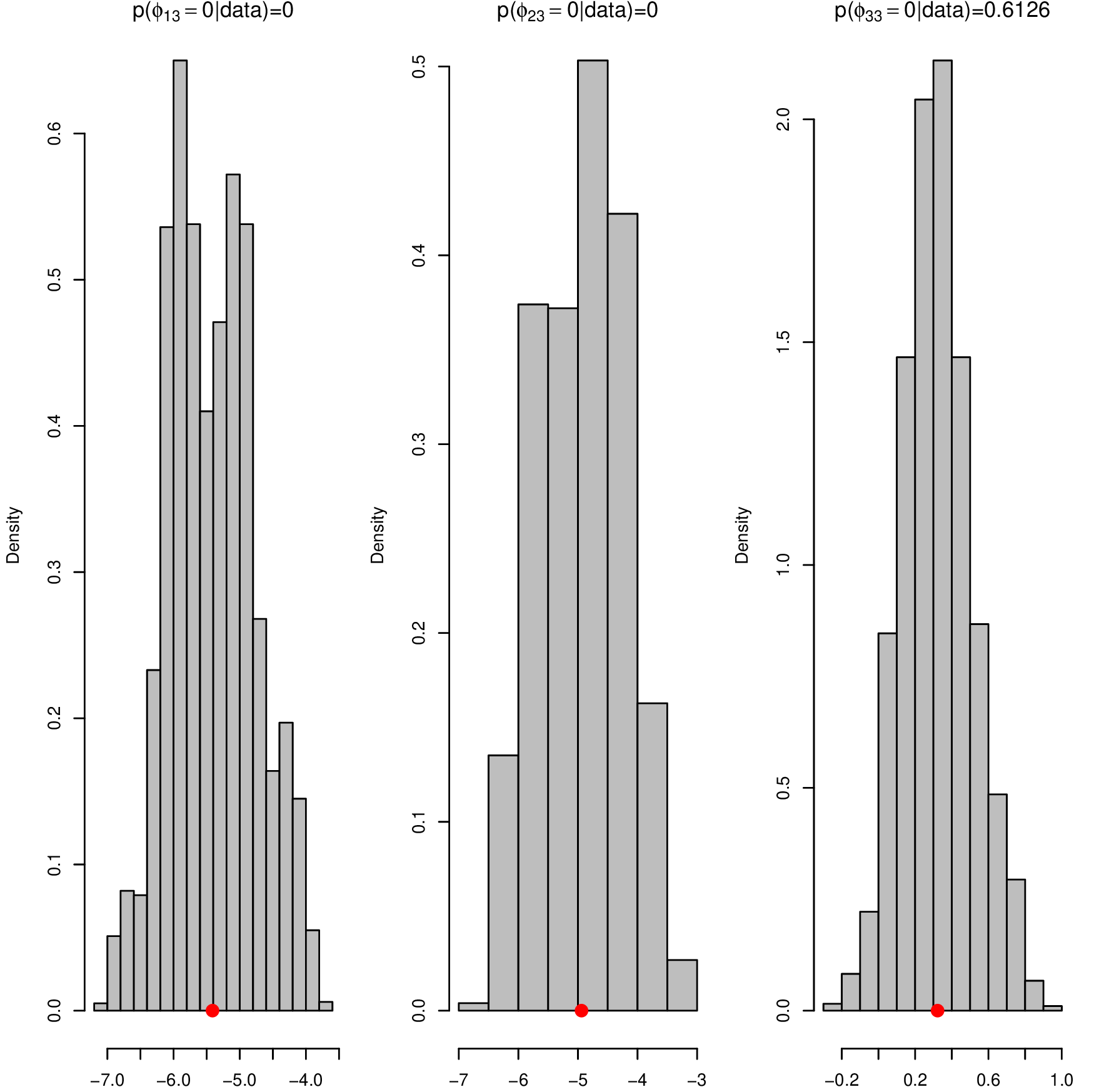}
\caption{Posterior distribution of the connectivity regions $\phi_{ij}$ for the fMRI application. The dots represent the posterior mean of the connectivity regions.}
\label{fig:phisapply}
\end{center}
\end{figure}

\begin{figure}
\begin{center}
\includegraphics[scale=0.5,keepaspectratio]{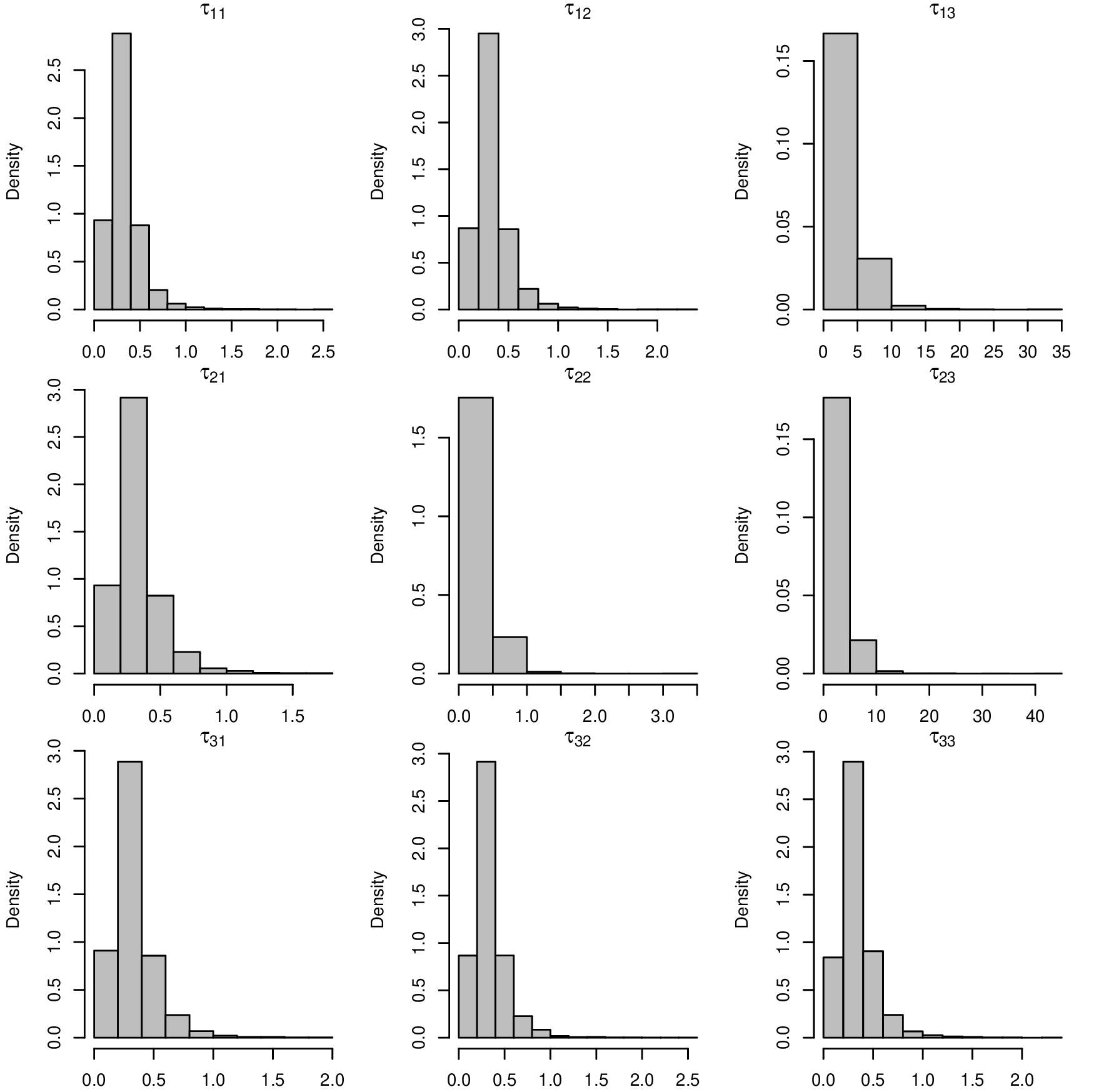}
\caption{Posterior precisions of the point-mass prior over the connectivity regions.}
\label{fig:precapply}
\end{center}
\begin{center}
\includegraphics[scale=0.5, keepaspectratio]{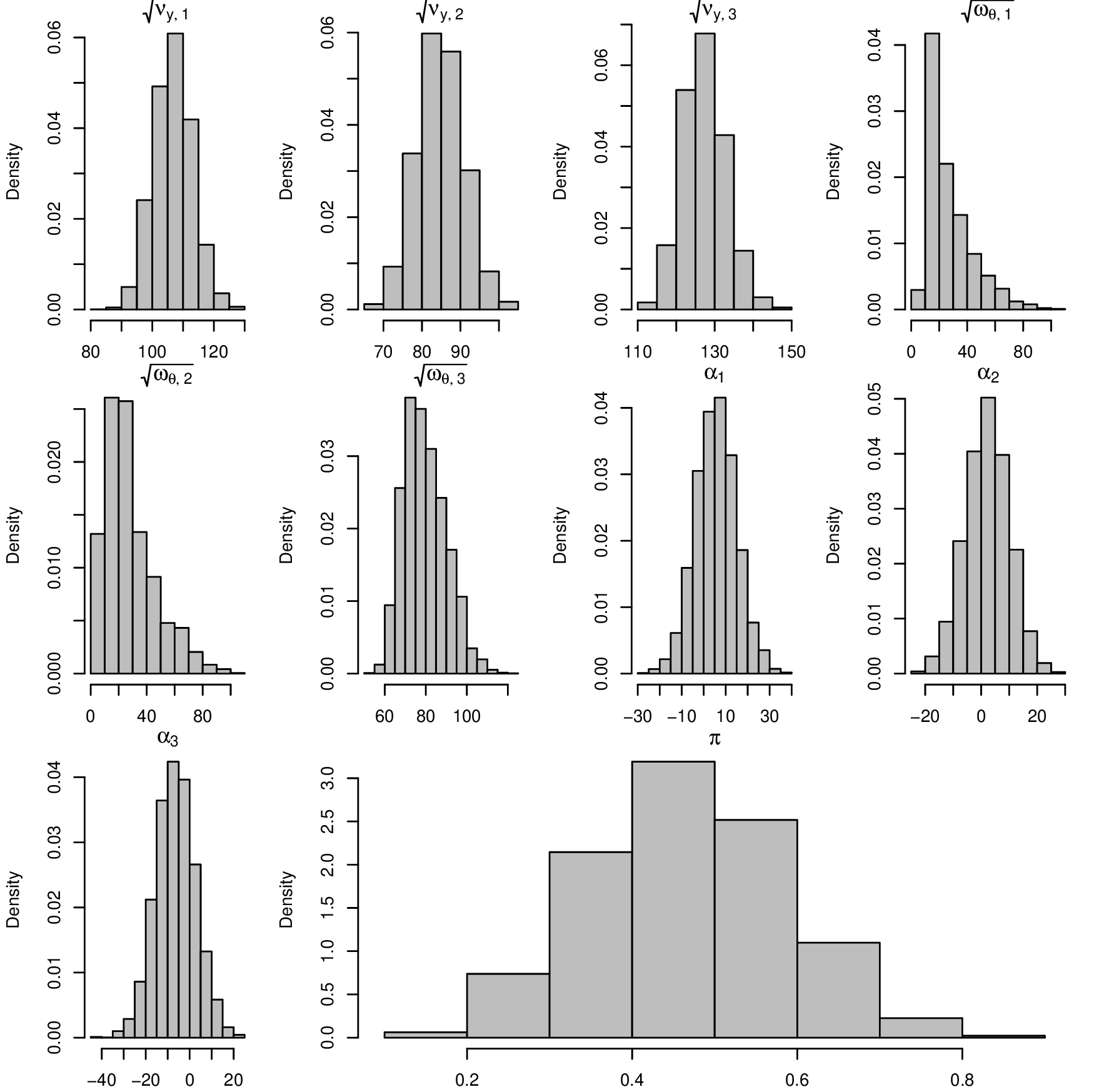}
\caption{Posterior distributions: observational and state standard deviations, trends and weights for the fMRI application.}
\label{fig:varapply}
\end{center}

\end{figure}

Table 3 shows the posterior summary for the connectivity parameters. Figures \ref{fig:phisapply} to \ref{fig:precapply} display the results obtained using the proposed Bayesian
approach. Our approach indicates that the probability of the regions DLFCP and LG or DLFCP and MG being connected is high ($P(\phi_{13}\neq
0|\text{data})=P(\phi_{23}\neq 0|\text{data})=1$). Also, with probability equal to 0.61, the posterior mean of $\phi_{33}$ is different from zero.
Therefore, there is evidence of a positive self-feedback at DLFCP. On the other hand, there was not self-feedback in the two sites of control, LG and MOG,
($P(\phi_{11}=0|\text{data})=1$ and $P(\phi_{22}=0|\text{data})=0.99$). Because of the posterior probability $P(\phi_{31}=0|\text{data})=0.99$ and
$P(\phi_{32}=0|\text{data})=0.98$, we cannot conclude that there is any influence on the MOG from the LG and DLFCP regions. Our results showed that
there was not substantial suppression from MOG on LG ($P(\phi_{12}=0|\text{data})=0.99$) and also from LG on MOG ($P(\phi_{21}=0|\text{data})=1$). The results are
consistent with \citeasnoun{Banish}, and the connectivity between the regions is consistent with the theory of attentional control.

\section{Discussion} \label{sec:conclusions}

To model the connectivity between brain signals for a particular subject, we propose a multivariate dynamic Bayesian model that addresses the main limitations of previous approaches to this problem. The introduction of a point-mass prior for the connectivity parameters allows us to perform automatic model selection over the set of all possible models. Our proposal also includes a new weakly informative default
variance state prior that is suitable for modelling the high frequency behavior characteristic of fMRI data. This prior induces robustness and shrinkage for the sparse state signals leading to more
coherent inference for the connectivity parameters. We showed that the proposed model
works in a large number of distinct scenarios where different signal/noise ratio values are considered. Finally, when the proposed approach was applied to fMRI data for a particular subject for
static connectivity parameters over time, we obtained accurate results in accordance with the theory of attentional control.\\

\textbf{Acknowledgements}\\

We thank Moon-Ho Ringo Ho for his help in the preparation of the fMRI data.

\clearpage

\bibliographystyle{agsm}
\citationstyle{agsm}
\bibliography{biblio_pF}

\end{document}